\documentclass{article}

\usepackage{arxiv}

\usepackage[utf8]{inputenc} 
\usepackage[T1]{fontenc}    
\usepackage{hyperref}       
\usepackage{url}            
\usepackage{booktabs}       
\usepackage{amsfonts}       
\usepackage{nicefrac}       
\usepackage{microtype}      
\usepackage{lipsum}

\usepackage{morefloats}
\usepackage{color}
\usepackage{multirow}
\usepackage{latexsym}
\usepackage{amsmath,amssymb,amsthm,graphicx}
\usepackage{dsfont}
\usepackage{enumitem}
\usepackage{hyperref}
\usepackage{arydshln}
\usepackage{lscape}
\usepackage{adjustbox}
\newtheoremstyle{thm}
{9pt}
{9pt}
{\itshape}
{}
{\bfseries}
{.}
{ }
{}
\theoremstyle{thm}
\DeclareMathAlphabet{\mathscr}{OT1}{pzc}{m}{it}
\newtheorem{theorem}{Theorem}[section]

\newtheorem{corollary}[theorem]{Corollary}

\newtheoremstyle{def}
{9pt}
{9pt}
{}
{}
{\bfseries}
{.}
{ }
{}
\theoremstyle{def}

\newtheorem{example}[theorem]{Example}

\newcommand{\R}{\mathbb{R}} 
\newcommand{\E}{\mathbb{E}} 
\newcommand{\PP}{\mathbb{P}} 

\newcommand{\fse}{\overset{a.s.}{\longrightarrow}}
    \def\cd{\stackrel{\mathcal{D}}{\longrightarrow}}

\renewcommand{\footnoterule}{%
	\kern -3.5pt
	\hrule width \textwidth height 1pt
	\kern 3.5pt
}

\makeatletter
\def\blfootnote{\xdef\@thefnmark{}\@footnotetext}
\makeatother

\title{A new omnibus test of fit based on a characterisation of the uniform distribution}

\author{B. Ebner\\
 Institute of Stochastics, \\
Karlsruhe Institute of Technology (KIT), \\
Englerstr. 2, 76133 Karlsruhe,\\ Germany. \\
\href{mailto:Bruno.Ebner@kit.edu}{Bruno.Ebner@kit.edu}\\
\And
S. C. Liebenberg\\
School of Mathematical and Statistical Sciences,\\ North-West University,\\ South Africa. \\
\href{mailto:Shawn.Liebenberg@nwu.ac.za}{Shawn.Liebenberg@nwu.ac.za}
\And I. J. H. Visagie\\
School of Mathematical and Statistical Sciences,\\ North-West University,\\ South Africa.\\
\href{mailto:Jaco.Visagie@nwu.ac.za}{Jaco.Visagie@nwu.ac.za}\\
}

\begin{document}

\date{\today}
\maketitle


\begin{abstract}
In this paper, we revisit the classical goodness-of-fit problems for univariate distributions; we propose a new testing procedure based on a characterisation of the uniform distribution. Asymptotic theory for the simple hypothesis case is provided in a Hilbert-Space setting, including the asymptotic null distribution as well as values for the first four cumulants of this distribution, which are used to fit a Pearson system of distributions as an approximation to the limit distribution. Numerical results indicate that the null distribution of the test converges quickly to its asymptotic distribution, making the critical values obtained using the Pearson system particularly useful. Consistency of the test is shown against any fixed alternative distribution and we derive the limiting behaviour under fixed alternatives with an application to power approximation. We demonstrate the applicability of the newly proposed test when testing composite hypotheses. A Monte Carlo power study compares the finite sample power performance of the newly proposed test to existing omnibus tests in both the simple and composite hypothesis settings. This power study includes results related to testing for the uniform, normal and Pareto distributions. The empirical results obtained indicate that the test is competitive. An application of the newly proposed test in financial modelling is also included.
\end{abstract}

\section{Introduction}
\label{Intro}

In this paper, we revisit the classical goodness-of-fit problems of testing a simple as well as a composite hypothesis and propose a new testing procedure based on a characterisation of the uniform distribution that is distribution-free when testing a simple hypothesis. To be precise let $X$ be a random variable with continuous distribution function $F$ and let $X_1,\ldots,X_n$ be independent and identically distributed (i.i.d.) copies of $X$. The problem of interest is testing
\begin{equation}\label{eq:allgH0}
H_0:\;F=F_0,
\end{equation}
where $F_0$ is some specified continuous distribution, against the general alternative $H_1:\, F\not=F_0$. Define $U=F_0(X)$ and $U_j=F_0(X_j),j=1,\ldots,n$, respectively, then  invoking the probability integral transform  the hypothesis in (\ref{eq:allgH0}) is restated as
\begin{equation}\label{eq:H0unif}
   H_0:\, U\sim{\cal U}(0,1),
\end{equation}
where ${\cal U}(0,1)$ stands for the uniform law on the unit interval. In the simplest case, we are interested in testing the hypothesis that the transformed observed data $U_1,\ldots,U_n$ are realised from a standard uniform distribution, an assumption which is of interest, for instance, when testing pseudo random number generators; see \cite{Knu2011}.


The classical goodness-of-fit problem has been extensively studied in the literature and there is ongoing research interest into this problem; for a recent reference, see \cite{BGT:2018}. Classical procedures for testing the hypothesis in (\ref{eq:allgH0}) are based on the empirical distribution function. These procedures include the prominent Kolmogorov-Smirnov test, the Cram\'{e}r-von Mises test, the Anderson-Darling test as well as the Watson test; see \cite{DS1986}. It is interesting to note that these classical procedures are distribution free in the sense that the distribution of the test statistic under the null hypothesis does not depend on the specified form of $F_0$. This fact is typically proved by transforming the random variable to uniformity under the null hypothesis using the inverse of the probability integral transform, see section 19 in \cite{V:1998}. Other methods for testing the simple hypothesis \eqref{eq:allgH0} are based on spacings, see \cite{qu1977,sh1950}, on the integrated empirical process, see \cite{HN:2000}, on the likelihood ratio, see \cite{zh2002}, on order statistics, see \cite{fr1987book}, on maximum correlation, see \cite{FG:2003} and on a  characterisation of the uniform distribution, see \cite{HS:1993,B:2020,vol2020}.

Below, we propose a testing procedure for (\ref{eq:H0unif}) based on a characterisation of the uniform distribution obtained as a special case of Theorem 1 of \cite{A:1991}, which adapted to the uniform distribution reads as follows.
\begin{theorem}\label{thm:char}
Let $U$ be a random variable taking values in $(0,1)$. Then $U\sim {\cal U}(0,1)$ if, and only if,
\begin{equation*}
\E(U|U\ge t)=\frac{1}{2}\left(1+\frac{t(1-t)}{\mathbb{P}(U\ge t)}\right),\quad 0<t<1.
\end{equation*}
\end{theorem}
Simple calculations allow us to rewrite the characterisation in Theorem \ref{thm:char} in the following form which lends itself to a simple formulation of the proposed test statistic.
\begin{corollary}\label{cor:char}
Let $U$ be a random variable taking values in $(0,1)$. Then $U\sim {\cal U}(0,1)$ if, and only if,
\begin{equation}\label{eq:char2}
\E\left((2U-1)\mathbf{1}\{U\ge t\}\right)-t(1-t)=0,\quad 0<t<1,
\end{equation}
where $\mathbf{1}\{\cdot\}$ is the indicator function.
\end{corollary}

The newly proposed test of fit is based on Corollary \ref{cor:char}, hence a characterisation of the uniform distribution. The idea of proposing tests motivated by characterisations goes back to \cite{Lin1953} and are known to have desirable properties, see \cite{Nik2017} for a discussion of these types of tests and their efficiencies. Replacing the expected value in \eqref{eq:char2} by its empirical counterpart, we propose the $L^2$-type statistic
\begin{equation}\label{eq:Un}
T_{n}=n\int_{0}^1\left|\frac1n\sum_{j=1}^n\left(2U_j-1\right)\mathbf{1}\{U_j\ge t\}-t(1-t)\right|^2\mbox{d}t.
\end{equation}

The test statistic in (\ref{eq:Un}) can be expressed in the following simply calculable form
\begin{equation}\label{eq:expl}
T_{n}=\frac1n\sum_{i,j=1}^n\left(4U_jU_k-2(U_j+U_k)+1\right)\min(U_j,U_k)-\frac{1}{3}\sum_{j=1}^n\left(2U_j-1\right)U_j^2(3-2U_j)+\frac{n}{30}.
\end{equation}
Since the integrand in \eqref{eq:Un} should be close to 0 for all $t\in(0,1)$ under uniformity, $H_0$ in (\ref{eq:H0unif}) is rejected for large values of $T_n$, where large is to be made precise in Section \ref{sec:Asyprop}.

The testing problem of uniformity is of ubiquitous interest since it can be used to test the hypothesis that observed data are realised from a parametric family of distributions indexed by a (possibly vector valued) unknown parameter, $\theta$. Denote such a hypothesised family of distribution functions by $\mathscr{F}_\Theta=\{F(\cdot,\theta):\,\theta\in\Theta\}$, where $\Theta$ is the parameter space of the family of distributions. The current paper demonstrates how the newly proposed test can be amended to test this composite hypothesis. The main idea is to calculate some estimate of $\theta$, say $\widehat{\theta}$, and to calculate $U_j=F_{\widehat{\theta}}(X_j),j=1,\ldots,n$. The test statistic in (\ref{eq:expl}) is then calculated based on $U_1,\ldots,U_n$. Testing a composite hypothesis is, of course, complicated by the fact that the distribution function is not completely specified under $H_0$ and includes one or more parameters which are required to be estimated. As a result, $U_j$ and $U_k$ are no longer independent when $j \neq k$. In the case of location scale families of distributions, relatively simple modifications to the testing procedures allow us to test the required hypotheses; see, for example \cite{EHS2020} and \cite{HV2020}. When dealing with distributions including one or more shape parameters, the situation is complicated further. For the procedure to perform the test in this case, the interested reader is referred to \cite{ABE2019} as well as \cite{EL2021}.

The remainder of the paper is organised as follows. In Section \ref{sec:Asyprop} we consider the simple hypothesis case and we derive the limit null distribution of $T_n$. We also derive the first four cumulants of the limit distribution in order to fit a Pearson system. Furthermore, this section shows omnibus consistency against all alternatives and examines the behaviour of the test under fixed alternatives. In Section \ref{sec:compHyp} we extend the problem to test the composite hypothesis of fit to families of distribution. Special attention is paid to the normal and Pareto distributions. Section \ref{sec:SimuUn} provides a comparative Monte Carlo simulation study containing critical values and empirical powers in the case of both simple and composite null hypotheses. Section 5 shows an application of the newly proposed test found in financial modelling. The paper concludes in Section 6 with an overview of open research problems.

\section{Testing simple hypotheses}\label{sec:Asyprop}

In this section, we develop the theory of the newly proposed test for testing simple hypotheses. Due to the $L^2$-structure of the test statistic, a convenient setting to derive asymptotic results is the separable Hilbert space $\mathbb{H}=L^2((0,1),\mathcal{B},{\rm d}t)$ of (equivalence classes of) measurable functions $f:(0,1) \rightarrow \R$ satisfying $\int_0^1 |f(t)|^2 \, {\rm d}t < \infty$. Here, $\mathcal{B}=\mathcal{B}_{|(0,1)}$ denotes the Borel sigma-field restricted to $(0,1)$. The scalar product and the norm in $\mathbb{H}$ is denoted by
\begin{equation*}
\langle f,g \rangle_{\mathbb{H}} = \int_0^1 f(t)g(t)\,{\rm d}t, \quad \|f\|_{\mathbb{H}} = \langle f,f \rangle_{\mathbb{H}}^{1/2}, \quad f,g \in \mathbb{H},
\end{equation*}
respectively. Note that with
\begin{equation*}
\widehat{Z}_n(t)=\frac{1}{\sqrt{n}}\sum_{j=1}^n\left(2U_j-1\right)\mathbf{1}\{U_j\ge t\}-t(1-t),\quad 0< t <1,
\end{equation*}
we have $T_n=\|\widehat{Z}_n\|_{\mathbb{H}}^2$ and hence a direct application of the central limit theorem in Hilbert spaces proves the first part of the following theorem and an application of the continuous mapping theorem shows the second part. In the following we denote by $\cd$ weak convergence of random elements (or variables).

\begin{theorem}\label{thm:asydistsh}
Let $U\sim\mathcal{U}(0,1)$ and let $U_1,\ldots,U_n,\ldots$ be i.i.d. copies of $U$.
\begin{enumerate}
\item[a)] There exists a centred Gaussian element $Z$ of $\mathbb{H}$ with covariance kernel
\begin{equation*}
K_{Z}(s,t)=\frac{1-(2\max(s,t)-1)^3}{6}-st(1-s)(1-t),\quad s,t\in(0,1),
\end{equation*}
such that $\widehat{Z}_n\cd Z$ in $\mathbb{H}$ as $n\rightarrow\infty$.
\item[b)] $T_n\cd T_{\infty}=\|Z\|_{\mathbb{H}}^2$ as $n\rightarrow\infty$.
\end{enumerate}
\end{theorem}
Note that by direct evaluation of integrals the first two cumulants of the distribution of $U_\infty$ are
\begin{equation*}
\kappa_1=\E(T_\infty)=\int_0^1K_{Z}(t,t)\,\mbox{d}t=\frac2{15}\quad\mbox{and}\quad \kappa_2=\mathbb{V}(T_\infty)=2\int_0^1\int_0^1K_{Z}^2(s,t)\,\mbox{d}s\mbox{d}t=\frac{109}{4050},
\end{equation*}
where $\mathbb{V}(\cdot)$ is used to denote the variance.
Following the methodology in \cite{H:1990,S:1976}, we calculate the third and fourth cumulants by
\begin{equation*}
\kappa_j=2^{j-1}(j-1)!\int_{0}^1K_j(t,t)\,\mbox{d}t,
\end{equation*}
where $K_j(s,t)$, the $j^{\text{th}}$ iterate of $K_{Z}(s,t)$, is given by
\begin{eqnarray*}
K_{j}(s,t)&=&\int_0^1K_{j-1}(s,u)K_{Z}(u,t)\,\mbox{d}u,\quad j\ge2,\\
K_1(s,t)&=&K_{Z}(s,t).
\end{eqnarray*}
Direct calculation shows that $\kappa_3=502883/40540500$ and $\kappa_4=200311667/23260111875$. These results can be used in packages that implement the Pearson system directly; see \cite{JKB:1994}, Chapter 12, Section 4.1. In the statistical computing language \texttt{R}, see \cite{CRAN}, we use the package \texttt{PearsonDS}, see \cite{BK:2017}, to approximate the asymptotic critical values of $T_n$. In Section \ref{sec:SimuUn}, we provide critical values obtained using Monte Carlo simulation and we compare these values to the approximate critical values obtained using the Pearson system. The critical values obtained using the Pearson system is close to those obtained using Monte Carlo simulation, even for relatively small sample sizes; see Table \ref{CritvalsUnif}.

In view of consistency of the test, we have, as a consequence of the strong law of large numbers in Hilbert spaces, that
\begin{equation}\label{eq:aslimit}
\frac{T_n}{n}\fse\Delta=\int_0^1\left|2\E\left(U\mathbf{1}\{U\ge t\}\right)-\mathbb{P}(U\ge t)-t(1-t)\right|^2\,{\rm d}t,
\end{equation}
here $\fse$ stands for almost sure convergence. Since $\Delta=0$ holds if and only if $U\sim\mathcal{U}(0,1)$ by Corollary \ref{cor:char} and $\Delta>0$ otherwise, the test based on $T_n$ is consistent against any fixed alternative.

In the following paragraph, we use Theorem 1 of \cite{BEH:17} to show the behaviour of $T_n$ under fixed alternatives. In this spirit, we assume that the distribution of the $(0,1)$-valued random variable $U$ is absolutely continuous. To make comparison easy, we use the notation of \cite{BEH:17}. Define $\Psi(t)=\E\left((2U-1)\mathbf{1}\{U\ge t\}\right)$, and write $z(t)=\Psi(t)-t(1-t)$, $t\in(0,1)$, such that $\Delta=\|z\|^2_{\mathbb{H}}$. With $Z_n(\cdot)=\frac1{\sqrt{n}}\widehat{Z}_n(\cdot)$, we thus have
\begin{equation}\label{eq:deco}
    \sqrt{n}\left(\frac{T_n}{n}-\Delta\right)=\sqrt{n}\left(\|Z_n\|_{\mathbb{H}}^2-\|z\|^2_{\mathbb{H}}\right)=\sqrt{n}\langle Z_n-z,Z_n+z\rangle_{\mathbb{H}}=2\langle\sqrt{n}(Z_n-z),z\rangle_{\mathbb{H}}+\frac1{\sqrt{n}}\|\sqrt{n}(Z_n-z)\|_{\mathbb{H}}^2.
\end{equation}
We write
\begin{equation*}
    W_n(t)=\sqrt{n}\left(Z_n(t)-z(t)\right)=\sqrt{n}\left(\frac1n\sum_{j=1}^n(2U_j-1)\mathbf{1}\{U_j\ge t\}-\Psi(t)\right),\quad t\in(0,1),
\end{equation*}
and note that $W_n$ is a centred sum of i.i.d. random elements of $\mathbb{H}$. An application of the central limit theorem in Hilbert spaces shows the following statement.
\begin{theorem}\label{thm:GPFA}
Let $U$ be a $(0,1)$ valued random variable with absolutely continuous distribution and let $U_1,\ldots,U_n$ be i.i.d. copies of $U$. There exists a centred Gaussian element $W$ of $\mathbb{H}$ with covariance kernel
\begin{equation*}
K_{W}(s,t)=\E\left(\left(2U-1\right)^2\mathbf{1}\{U\ge\max(s,t)\}\right)-\Psi(s)\Psi(t),\quad s,t\in(0,1),
\end{equation*}
such that $W_n\cd W$ in $\mathbb{H}$ as $n\rightarrow\infty$.
\end{theorem}
In Theorem \ref{thm:GPFA} we see by the continuous mapping theorem, that the asymptotic behaviour of the right hand side of \eqref{eq:deco} is completely determined by the weak limit of $W_n$. Thus, an application of Theorem 1 in \cite{BEH:17} directly proves the next Corollary.
\begin{corollary}\label{cor:FAL}
Under the assumptions of Theorem \ref{thm:GPFA} we have
\begin{equation*}
    \sqrt{n}\left(\frac{T_n}{n}-\Delta\right)\cd \mbox{N}(0,\sigma^2)
\end{equation*}
as $n\rightarrow\infty$, where $
\sigma^2=4\int_0^1\int_0^1K_W(s,t)z(s)z(t)\,\mbox{d}s\mbox{d}t$.
\end{corollary}
Below we show for several alternative distributions on $(0,1)$ that Corollary \ref{cor:FAL} fully determines the behaviour of $T_n$. Let $\beta(a,b)$, $a,b>0,$ denote the beta distribution with probability density function$
    f(t)=(B(a,b))^{-1}t^{a-1}(1-t)^{b-1},\; 0< t< 1$,
where $B(\cdot,\cdot)$ is the beta function. The following results were partially derived with help of the computer algebra system Maple, see \cite{Maple2019}. For a short notation we write $s \vee t=\max(s,t)$, $s,t\in \R$.
\begin{example}\label{ex:Beta}
\begin{enumerate}
    \item[a)] Let $U\sim \beta(2,2)$. Direct evaluation of the integrals shows $\Delta=1/210\approx0.004761904762$, $\psi(t)=3t^2(1-t)^2$. In Corollary \ref{cor:FAL}, we have
    \begin{equation*}
     K_W(s,t)=\frac{24}5(s \vee t)^5-12(s \vee t)^4+10(s \vee t)^3-3(s \vee t)^2-9t^2(t - 1)^2s^2(s - 1)^2+\frac15,\quad 0< s,t< 1.
    \end{equation*}
    With this formula we explicitly calculate $\sigma^2=107297/94594500\approx0.001134283706$.
    \item[b)] Let $U\sim \beta(2,3)$. In this asymmetric case, we have $\Delta=71/2310\approx0.03073593074$, $\psi(t)=24t^5/5 + 15t^4 - 16t^3 + 6t^2- 1/5 $, and 
    \begin{eqnarray*}
     K_W(s,t)&=&-8(s \vee t)^6 +\frac{144}{5}(s \vee t)^5-39(s \vee t)^4+24(s \vee t)^3-6(s \vee t)^2\\&&-\frac1{25}((24t^2 - 3t - 1)(t - 1)^3(24s^2 - 3s - 1)(s - 1)^3)+\frac15,\quad 0< s,t< 1,
    \end{eqnarray*}
    and $\sigma^2=13088573/2948195250\approx0.004439520415$.
    \item[c)] Let $U\sim \beta(1,1/2)$. Here, we have $\Delta=53/945\approx0.05608465608$, $\psi(t)=\frac13(2t+1)\sqrt{1-t} $.  The covariance kernel is
    \begin{equation*}
     K_W(s,t)=\frac1{15}\left(12(s \vee t)^2-4(s \vee t)+7\right)\sqrt{1-s \vee t}-\frac19(2s + 1)(2t + 1)\sqrt{(1 - s)(1-t)},\quad 0< s,t< 1,
    \end{equation*}
    and $\sigma^2=426456598/10854718875\approx0.03928766861$.
  \item[d)] Let $U\sim \beta(1/2,1/2)$, which corresponds to the arcsin distribution. Here, some integrals could not be computed explicitly and numerical integration methods where used. We have $\Delta\approx0.007130789095$, $\psi(t)=\frac2\pi\sqrt{t(1-t)} $, and 
    \begin{eqnarray*}
     K_W(s,t)&=&\frac2\pi(s \vee t)^{3/2}\sqrt{1-s \vee t}-\frac1\pi\sqrt{(s \vee t)(1-s \vee t)}-\frac1{2\pi}\arcsin\left(2(s \vee t)-1\right)
     \\&&-\frac4{\pi^2}\sqrt{s(1-s)t(1-t)}+\frac14,\quad 0< s,t< 1,
    \end{eqnarray*}
    and $\sigma^2\approx0.004386925128$.
\end{enumerate}
\end{example}
As we see in Example \ref{ex:Beta}, we can find explicit values for $\Delta$ and $\sigma^2$ in Corollary \ref{cor:FAL}, which clearly depend on the underlying alternative distribution $F$. This knowledge allows us to asymptotically approximate the power of the test that rejects $H_0$ if $T_n>c_n$, and $\lim_{n\rightarrow\infty}\PP_{H_0}(T_n>c_n)=\alpha$ by (11) in \cite{BEH:17}, namely
\begin{equation}\label{eq:powapp}
    \PP_F(T_n>c_n)\approx 1-\Phi\left(\frac{\sqrt{n}}{\sigma}\left(\frac{c_n}{n}-\Delta\right)\right).
\end{equation}

\begin{figure}[t]
    \centering
    \includegraphics[scale=0.2]{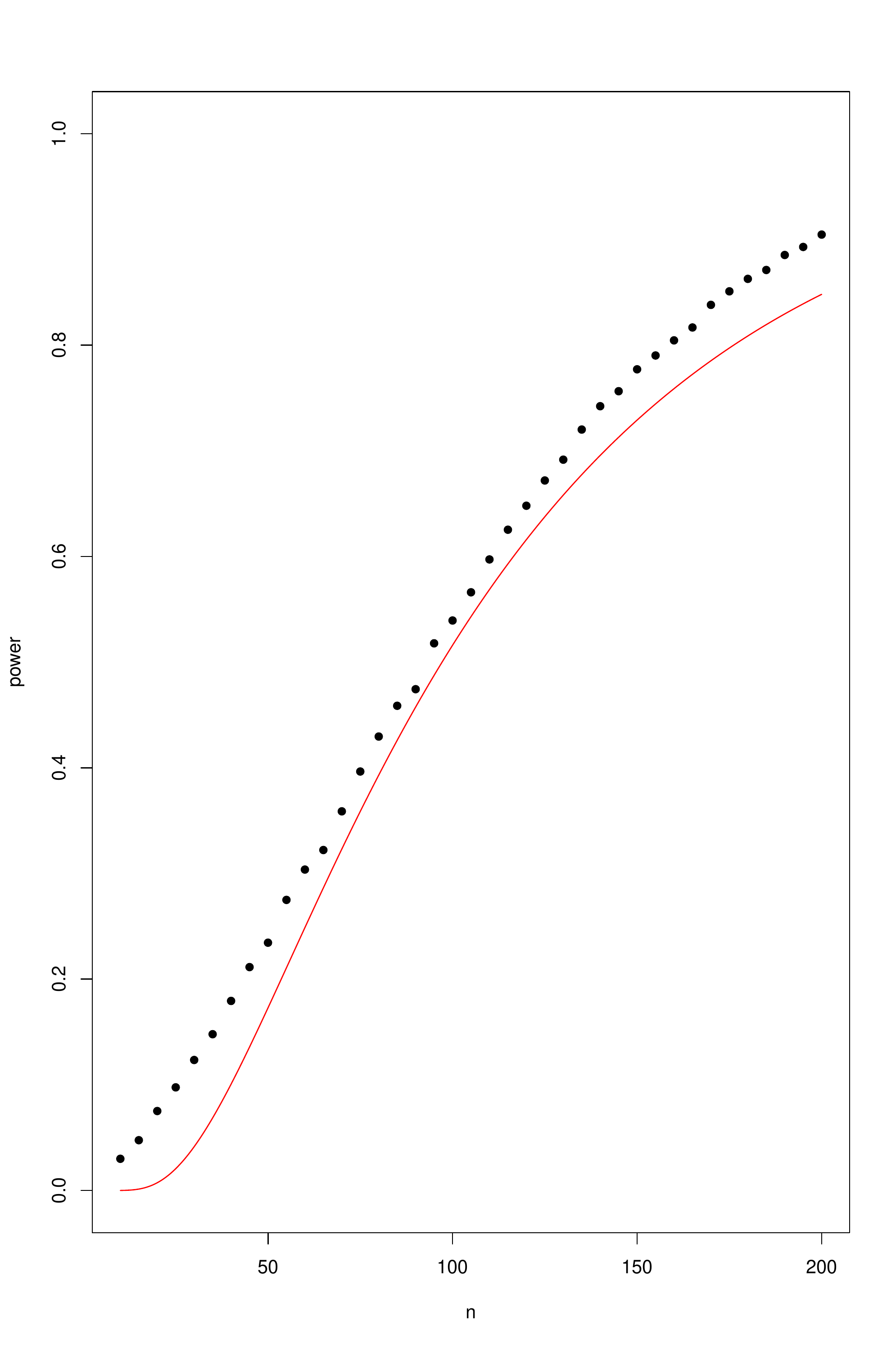}\includegraphics[scale=0.2]{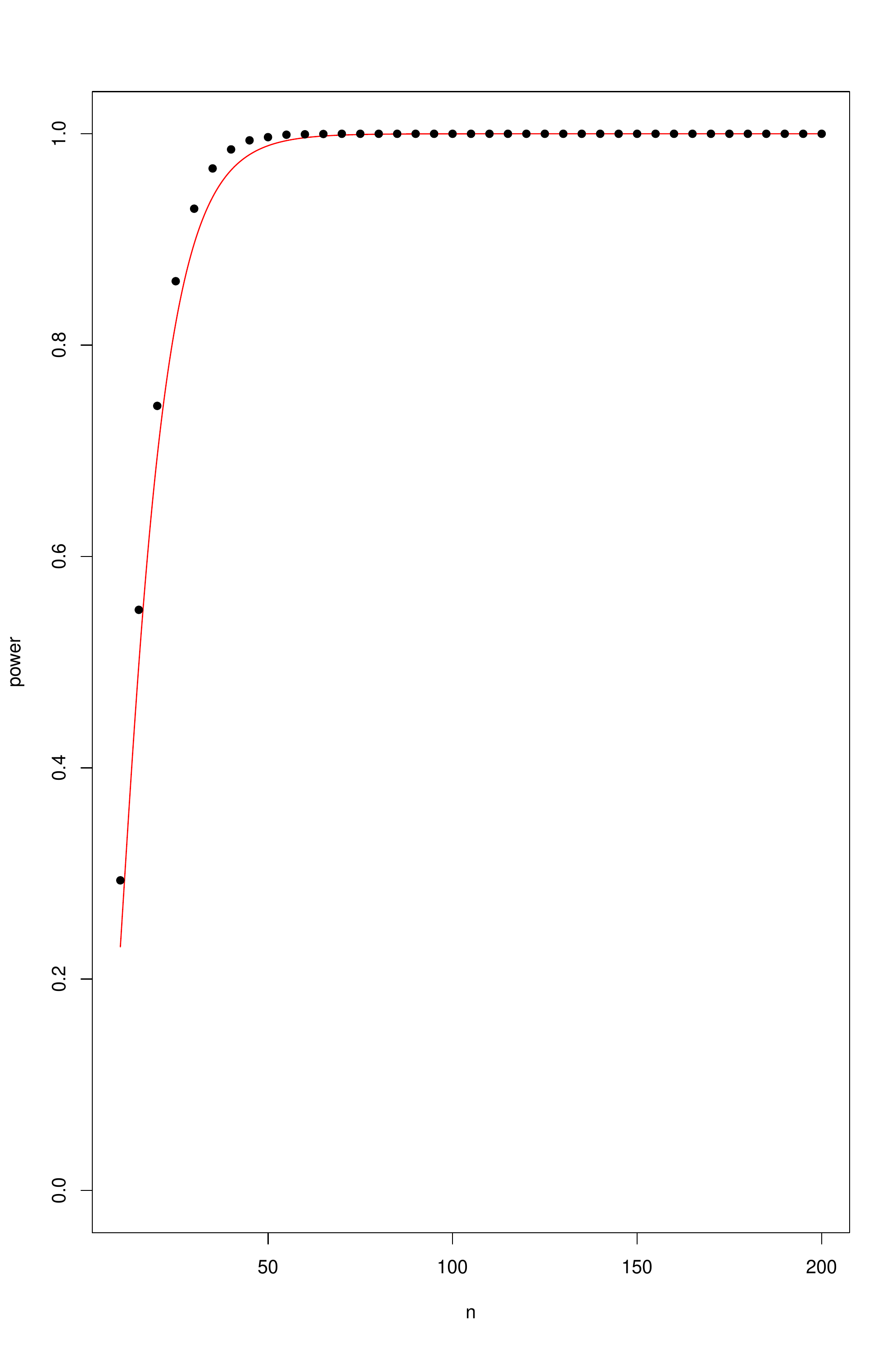}\includegraphics[scale=0.2]{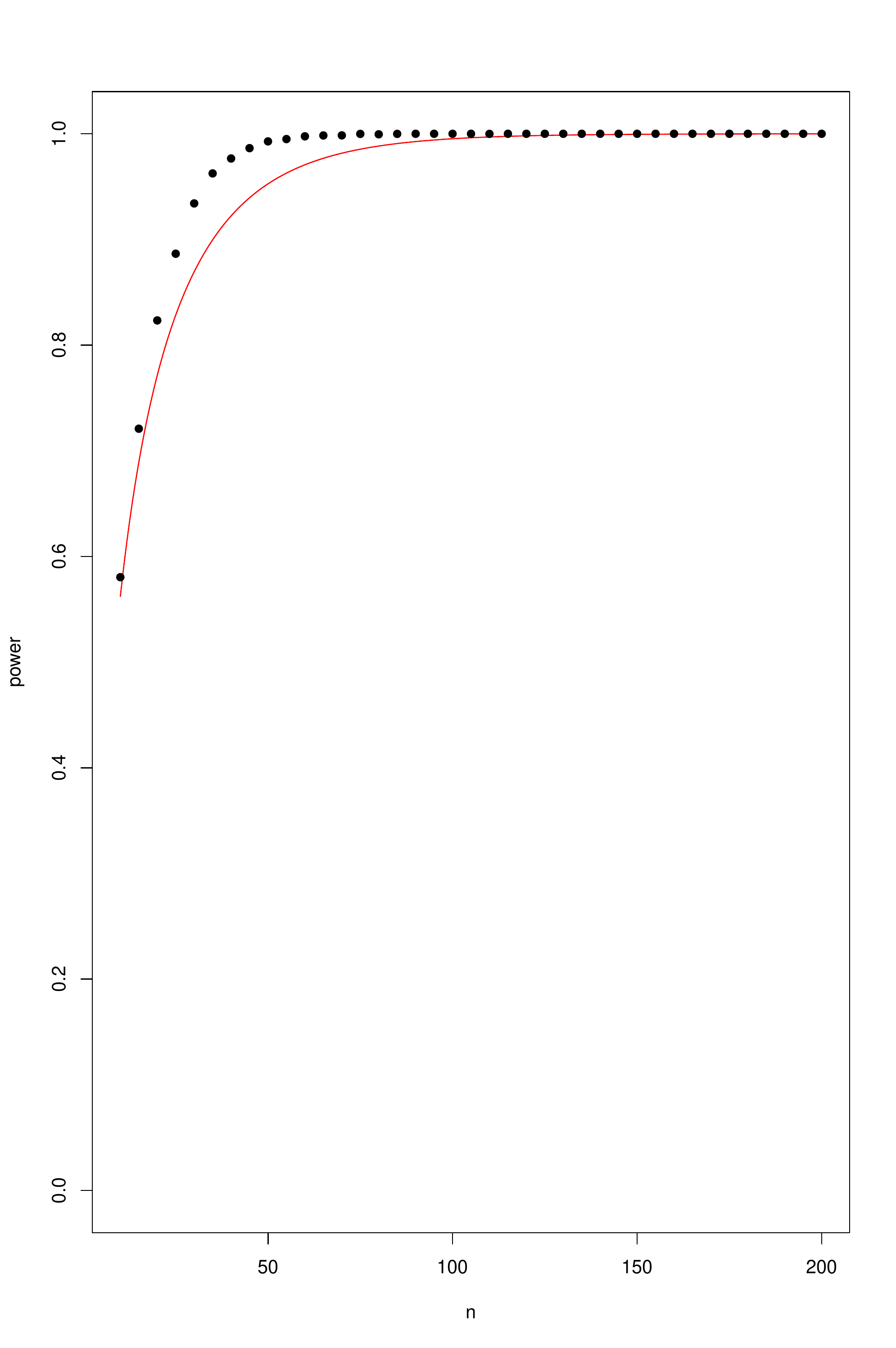}\includegraphics[scale=0.2]{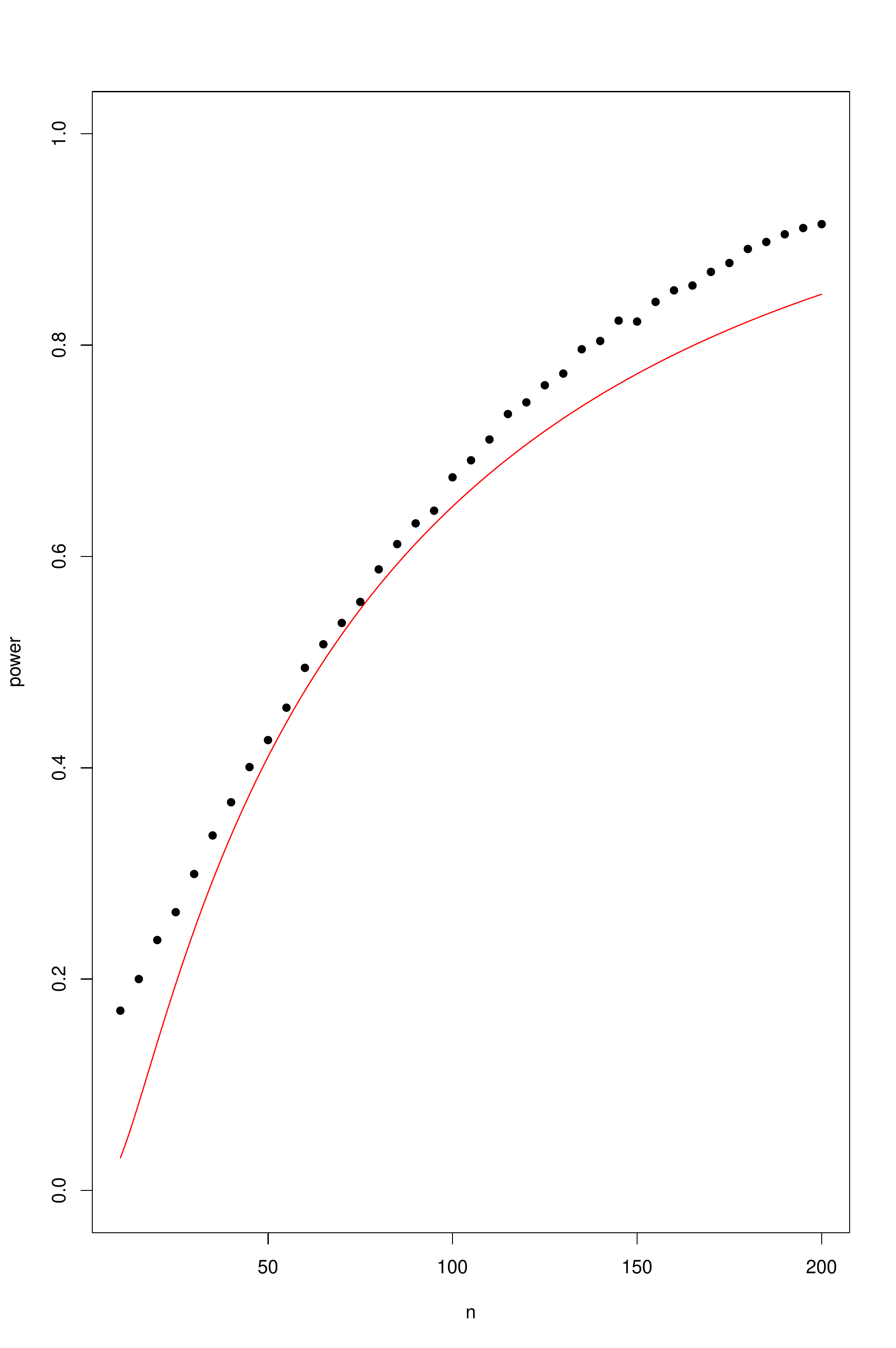}
    \caption{Power approximations and empirical powers of the test $T_n$ for the alternative distributions considered in Example \ref{ex:Beta}. From left to right, the alternatives are $\beta(2,2)$, $\beta(2,3)$, $\beta(1,1/2)$, and $\beta(1/2,1/2)$. The red line represents the power approximation by \eqref{eq:powapp} and each black dot represents a simulated empirical power for each sample size (obtained using 10 000 Monte Carlo replications).}
    \label{fig:powapp}
\end{figure}

In Figure \ref{fig:powapp} we compare the power approximation given by formula \eqref{eq:powapp} (in red) and an empirical power obtained using 10 000 replications (in black) for each alternative distribution considered in Example \ref{ex:Beta} for each of the sample sizes $n\in\{10,15,20,\ldots,200\}$. We fix the significance level at $\alpha=0.05$. The critical values used, $c_n$, were approximated by the Pearson system as described above. From these examples we conjecture that the power approximation in \eqref{eq:powapp} is a lower bound for the true power of the test, which is in accordance to the findings of \cite{BEH:17}; see Tables 2 and 7 therein. Further applications of results as in Example \ref{ex:Beta} are the calculation of confidence intervals for $\Delta$ or neighborhood-of-model validation; for details see Section 3 in \cite{BEH:17}.

\section{Testing composite hypotheses}\label{sec:compHyp}


Although developing the asymptotic results for the newly proposed statistic when testing composite hypotheses is beyond the scope of the current paper, we discuss how this test can be applied in this setting below. The setting for the testing problem of fit to parametric families of distributions is as follows. Let $X_1,\ldots,X_n$ be a sequence of i.i.d. random variables with distribution function $F$. Let $\mathscr{F}_\Theta=\{F(\cdot,\theta):\,\theta\in\Theta\}$ be a hypothesized parametric family of distributions. A classical problem of statistical inference is to test the null hypothesis
\begin{align}\label{eq:H0gen}
H_0:\;F\in\mathscr{F}_\Theta,
\end{align}
on the basis of $X_1,\ldots,X_n$ against general alternatives. This framework includes several classical goodness-of-fit problems such as testing for the normal and Pareto distributions, both of which are considered as special cases below. We propose new tests for these distributions by combining the uniformity test with the probability integral transform. The random variables $F(X_1,\theta),\ldots,F(X_n,\theta)$ are i.i.d. $\mathcal{U}(0,1)$ distributed. Let $\widehat{\theta}=\widehat{\theta}(X_1,\ldots,X_n)$ denote a consistent estimator for $\theta$. If the hypothesis in (\ref{eq:H0gen}) is true, then $F(X_1,\widehat{\theta}),\ldots,F(X_n,\widehat{\theta})$ will be approximately i.i.d. standard uniform random variables, at least for large $n$. However, the derivation of the asymptotic properties of the resulting test is substantially complicated by the dependency between the transformed random variables.



A classical problem in goodness-of-fit testing is testing the hypothesis that observed data are normally distributed; see \cite{MSA2014}. That is, testing the hypothesis in (\ref{eq:H0gen}) with $\mathcal{F}=\{\mbox{N}(\mu,\sigma^2): \mu\in\R, \sigma > 0\}$, where $\mbox{N}(\mu,\sigma^2)$ represents the normal distribution with mean $\mu$ and variance $\sigma^2$; denote the distribution function by $\Phi$ for the standard normal case. For a recent review of tests for normality in the multivariate case (which is also applicable in the univariate case), see \cite{EH2020}. Note that the family $\mathcal{F}$ is a location-scale family which implies that it is closed under location-scale transformations; $X\sim\mbox{N}(\mu,\sigma^2)$ if, and only if, $ (X-\mu)/\sigma \sim\mbox{N}(0,1)$. This motivates the use of the so called scaled residuals; $Y_{n,j}=(X_j -\widehat{\mu}_n)/\widehat{\sigma}_n$, $j=1,\ldots,n$, where $\widehat{\mu}_n=\frac1 n\sum_{j=1}^n X_j$ and $\widehat{\sigma}^{2}={\frac {1}{n}}\sum _{j=1}^{n}\left(X_{j}-\widehat{\mu}_n\right)^{2}$. The test in (\ref{eq:Un}) is applied to $\Phi(Y_1),\ldots,\Phi(Y_n)$.



As a second example of a parametric family of distributions, consider $\mathcal{F}=\{\mbox{P}(\beta): \beta > 0\}$, where $\mbox{P}(\beta)$ is the Pareto distribution with shape parameter $\beta$, see \cite{Par1897}. The distribution function of the $\mbox{P}(\beta)$ distribution is
$F_{\beta}(x) = 1-x^{-\beta}, x \geq 1$.
Unlike the normal distribution, the Pareto distribution contains a shape parameter and this class of distributions is not closed under location or scale changes. Interestingly, we are able to obtain fixed critical values for testing the hypothesis in (\ref{eq:H0gen}) for the class of Pareto distributions in spite of the presence of the shape parameter. To illustrate the reason for this, consider the maximum likelihood estimator of $\beta$ based on the sample $X_1,\ldots,X_n$;
\begin{equation*}
    \widehat{\beta}(X_1,\ldots,X_n) = \frac{n}{\sum_{j=1}^n \log{X_j}}.
\end{equation*}
Setting $Y_j = X_j^{\widehat{\beta}(X_1,\ldots,X_n)},j=1,\ldots,n$ and noting that
\begin{equation*}
    \widehat{\beta}(Y_1,\ldots,Y_n)
    = \frac{n}{\sum_{j=1}^n \log{Y_j}}
    = \frac{n}{\widehat{\beta}(X_1,\ldots,X_n)\sum_{j=1}^n \log{X_j}}
    = 1,
\end{equation*}
provides motivation for basing the test in (\ref{eq:Un}) on the transformed values $F_{\widehat{\beta}}(Y_1),\ldots,F_{\widehat{\beta}}(Y_n)$. An in depth review of the Pareto distribution as well as its applications and generalisations can be found in \cite{Arn2015}, while a recent review of goodness-of-fit tests for this distribution is provided in \cite{CDN2019}. For a recent paper proposing a new goodness-of-fit test for the Pareto distribution based on a characterisation, see \cite{obradovic2015goodness}.

\section{Numerical results}\label{sec:SimuUn}

In this section, we compare the finite-sample power performance of the newly proposed test to those of several existing tests. We report empirical powers in both the simple and composite hypothesis settings. In the first setting, the simple null hypothesis of standard uniformity is tested. When turning our attention to composite hypotheses, we test the composite hypotheses that data are realised from normal and Pareto distributions respectively.

\subsection{Alternative tests}

The tests included in the power studies below consist of a combination of classical procedures based on the empirical distribution function and existing tests for the uniform distribution which do not rely on a user-defined tuning parameter (as is the case for the newly proposed test). In the cases where composite hypotheses are tested, the required parameters are estimated using the method of maximum likelihood and the required transformations are affected as described above. Each of the tests are then applied to the transformed data, $Y_1,\ldots,Y_n$, in turn.

Note that, in the case of composite hypotheses, we do not compare the powers of the newly proposed test to those of the most powerful tests available for the specific distribution in question; for instance, we do not include the Shapiro-Wilk test when testing for normality. Omnibus tests cannot reasonably be expected to outperform these types of distribution specific tests. However, the aim of the simulation study is to compare the finite-sample power performance of the newly proposed test to those of other omnibus tests not containing tuning parameters.

The classical empirical distribution function based tests included are the Kolmogorov-Smirnov ($KS_n$), Cram\'{e}r-von Mises ($CV_n$), Anderson-Darling ($AD_n$) and Watson ($WA_n$) tests; see \cite{DS1986}. These tests are well known and their computational forms are omitted. The remaining tests were originally developed for testing the hypothesis of uniformity and are as follows.

The Sherman test, see \cite{sh1950}, is based on spacings and has the computational form
    \begin{equation*}
    S_{n}=\frac{1}{2} \sum_{j=1}^{n+1}\left|X_{(j)}-X_{(j-1)}-\frac{1}{n+1}\right|.
    \end{equation*}
The Kuiper test, proposed in \cite{Kui1960}, is closely related to the $KS_n$ test;
     \begin{equation*}
     K_{n} =\max _{j \in \{1,\ldots,n\}} \left(\frac{j}{n}-X_{(j)}\right)+\max _{j \in \{1,\ldots,n\}}\left(X_{(j)}-\frac{j-1}{n}\right).
 \end{equation*}
The Quesenberry–Miller test, proposed in \cite{qu1977}, is another spacings-based test, the computational form of which is
    \begin{equation*}
    QM_{n}=\sum_{j=1}^{n+1}\left(X_{(j)}-X_{(j-1)}\right)^{2}+\sum_{j=1}^{n}\left(X_{(j)}-X_{(j-1)}\right)\left(X_{(j+1)}-X_{(j)}\right).
\end{equation*}
The Frosini-Revesz-Sarkad test, see \cite{fr1987book}, is based on the order statistics;
\begin{equation*}
    FR_{n}=\frac{1}{\sqrt{n}} \sum_{j=1}^{n}\left|X_{(j)}-\frac{j-0.5}{n}\right| .
\end{equation*}
In \cite{zh2002}, Zhang proposes a likelihood ratio based test statistic with computational form
\begin{equation*}
ZC_n = \sum_{j=1}^{n}\left[\log \left\{\frac{F_{0}\left(X_{(j)}\right)^{-1}-1}{\left(n-\frac{1}{2}\right) /\left(j-\frac{3}{4}\right)-1}\right\}\right]^{2}.
\end{equation*}
Each of the tests considered rejects the null hypothesis for large values of the test statistic. Software packages are available for the calculation of the tests mentioned above; see, for example, \cite{uniftest} and \cite{PoweR}.

\subsection{Simulation setting}

Below we report estimated critical values of $T_n$ for the uniform, normal and Pareto distributions in Tables \ref{CritvalsUnif}, \ref{CritvalsNorm} and \ref{CritValsPareto}, respectively. The estimated critical values are provided for selected sample sizes ranging from $n=10$ to $n=500$. Each reported critical value is based on one million Monte Carlo simulations. In the case of the uniform distribution, we are able to approximate the asymptotic critical values of the test using the Pearson system; these critical values are given in the column labelled "$n=\infty$". The critical values reported for $T_n$ converge remarkably quickly to their asymptotic values, especially in the case of the uniform distribution.

\begin{table}[!htbp] \centering 
  \caption{Critical values for $T_n$ when testing for the uniform distribution} 
  \label{CritvalsUnif} 
\begin{tabular}{@{\extracolsep{5pt}} lccccccccccc} 
\\[-1.8ex]\hline 
\hline \\[-1.8ex] 
$n$ & $10$ & $20$ & $30$ & $50$ & $75$ & $100$ & $150$ & $200$ & $300$ & $500$ & $\infty$\\
\hline \\[-1.8ex] 
$\alpha=10\%$ & $0.328$ & $0.329$ & $0.329$ & $0.331$ & $0.331$ & $0.331$ & $0.330$ & $0.331$ & $0.331$ & $0.331$ & $0.332$ \\ 
$\alpha=5\%$ & $0.453$ & $0.456$ & $0.458$ & $0.461$ & $0.461$ & $0.462$ & $0.461$ & $0.462$ & $0.463$ & $0.462$ & $0.462$ \\ 
$\alpha=1\%$ & $0.763$ & $0.772$ & $0.777$ & $0.779$ & $0.783$ & $0.783$ & $0.785$ & $0.783$ & $0.786$ & $0.784$ & $0.785$ \\ 
\hline \\[-1.8ex] 
\end{tabular} 
\end{table} 

We also report empirical powers for testing the hypotheses of the uniform, normal and Pareto distributions against a range of alternative distributions. In order to ease comparison the highest power against each alternative considered is printed in bold. Below, we list the alternative distributions considered; in each case, we provide the density function as well as the notation used in the power tables to follow.
\begin{itemize}
\item The beta distribution is denoted by $\beta(a,b)$ as above and has density 
$f(a,b)=(B(a,b))^{-1}x^{a-1}(1-x)^{b-1}, 0\le x\le 1.$
\item The truncated normal distribution, ${TN}(\mu, \sigma^2)$, with $\Phi(\cdot)$ denoting the standard normal distribution function has density $f(\mu, \sigma)=\exp \left(-\frac{1}{2}\left(\frac{x-\mu}{\sigma}\right)^{2}\right)/\left(\sigma \sqrt{2 \pi}\left(\Phi\left((1-\mu)/\sigma\right)-\Phi\left(-\mu/\sigma\right)\right)\right),$ $0\leq x \leq 1.$
\item The Kumaraswamy distribution, $K(a,b)$, with density $f(a, b) =  a b x^{{a-1}}(1-x^{a})^{{b-1}},\quad 0\le x\le 1.$
\item The family of distributions used in \cite{stephens1974} and \cite{marhuenda2005}, denoted by $S_{1}(k)$, $S_{2}(k)$, $S_{3}(k)$ respectively have the following densities;
\begin{equation*}
\begin{aligned}
&f_{1}(x)=k(1-x)^{k-1}, \quad 0 \leq x \leq 1, \\
&f_{2}(x)=\left\{\begin{array}{ll}
k(2x)^{k-1} & \text { if } 0 \leq x \leq \frac{1}{2}, \\
k(2(1-x))^{k-1} & \text { if } \frac{1}{2} < x \leq 1,
\end{array}\right. \\
&f_{3}(x)=\left\{\begin{array}{ll}
k(1-2x)^{k-1} & \text { if } 0 \leq x \leq \frac{1}{2}, \\
k(2x-1)^{k-1} & \text { if } \frac{1}{2} < x \leq 1,
\end{array}\right.
\end{aligned}
\end{equation*}
where the index of $f$ above corresponds to that of $S$.
\item The Weibull distribution, $W(\theta)$, with density $f(x) = \theta x^{\theta -1 }\exp(-x^\theta), x \geq 0.$
\item The gamma distribution, $\Gamma(\theta)$, has density
$f(x) = (\tilde{\Gamma} \left( \theta \right))^{-1} x^{\theta -1}\exp(-x), x \geq 0,$ where $\tilde{\Gamma}$ is the gamma function.
\item The skew normal distribution, $SN(\theta)$, has density function $f(x) = \sqrt{2/\pi}\textrm{exp}\left(x^2/2\right) \Phi(\theta x), x \in \mathbb{R}.$
\item The linear failure rate distribution, $LFR(\theta)$, with density $f(x) = \left( 1+\theta x\right) \exp \left(-x-\theta x^{2}/2\right), x\geq0.$
\item The exponential geometric distribution, $EG(\theta)$, has density
$f(x) = (1-\theta) \textrm{e}^{-x}(1-\theta\textrm{e}^{-x})^{-2}, x \geq 0.$
\item Student's $t$ distribution, $t(\theta)$, has density $
    f(x) = \tilde{\Gamma}\left( (\theta+1)/2 \right)\left( 1+x^2/\theta \right)^{-(\theta+1)/2}/\left(\sqrt{\theta \pi}\tilde{\Gamma} \left(\theta/2\right)\right)$, $x\in\mathbb{R}.$
\item The chi-square distribution, $\chi^2(\theta)$, with density $f(x) = {2^{-k/2}}x^{k/2-1}\textrm{e}^{-x/2}/\tilde{\Gamma}(k/2),\, x \geq 0.$
\item The half normal distribution, $HN(\theta)$, has density $f(x) = \sqrt{2/(\pi \theta^2)}\textrm{exp}\left( -x^2/(2\theta^2) \right),\, x \geq 0.$
\end{itemize}

In the power tables, we use the notation $\mathcal{U}$, $Z$ and $\mbox{P}$ to denote standard uniform, normal and Pareto distributions. Several mixture distributions are also included in the power study below. If $A$ and $B$ are distributions, the notation
\begin{equation*}
    pA+(1-p)B,
\end{equation*}
is used to denote a mixture distribution in which the probabilities of sampling from $A$ and $B$ are $p$ and $1-p$, respectively.

Two sample sizes are considered in each case; $n=30$ and $n=50$. The nominal significance level is fixed at $5\%$ throughout. Each estimated power is based on 50 000 Monte Carlo simulations. The results for the uniform distribution are reported in Tables \ref{Unif30} and \ref{Unif50}. The results for the normal distribution are shown in \ref{Normal30} and \ref{Normal50}, while the powers associated with the Pareto distribution are given in Tables \ref{Pareto30} and \ref{Pareto50}. Note that the support of the Pareto distribution is $(1,\infty)$. To account for this, the alternative distributions listed in Tables \ref{Pareto30} and \ref{Pareto50} with support $(0,\infty)$ are translated by $1$; for example, $\Gamma(1)$ denotes the result of adding $1$ to a random variable from a gamma distribution.



\begin{table}[t] \centering 
  \caption{Empirical powers associated with testing for the uniform distribution, $n=30$} 
  \label{Unif30} 
\begin{tabular}{@{\extracolsep{5pt}} lcccccccccc} 
\\[-1.8ex]\hline 
\hline \\[-1.8ex]
Distributions & $KS_n$ & $CV_n$ & $AD_n$ & $W_n$ & $S_n$ & $K_n$ & $Q_n$ & $F_n$ & $ZC_n$ & $T_n$ \\
\hline
$\mathcal{U}$ & $5$ & $5$ & $5$ & $5$ & $5$ & $5$ & $5$ & $5$ & $5$ & $5$ \\ 
$\beta(2,2)$ & $11$ & $9$ & $11$ & $\mathbf{58}$ & $13$ & $\mathbf{52}$ & $11$ & $10$ & $46$ & $13$ \\ 
$\beta(2,3)$ & $76$ & $77$ & $77$ & $91$ & $34$ & $88$ & $55$ & $74$ & $\mathbf{92}$ & $\mathbf{93}$ \\ 
$\beta(1,0.5)$ & $83$ & $87$ & $93$ & $61$ & $53$ & $64$ & $56$ & $88$ & $\mathbf{94}$ & $\mathbf{94}$ \\ 
$\beta(0.5,0.5)$ & $27$ & $28$ & $\mathbf{67}$ & $61$ & $46$ & $58$ & $50$ & $28$ & $\mathbf{83}$ & $30$ \\ 
$TN(0,0.5)$ & $72$ & $\mathbf{78}$ & $76$ & $52$ & $24$ & $52$ & $40$ & $\mathbf{78}$ & $68$ & $76$ \\ 
$TN(0,1)$ & $10$ & $\mathbf{11}$ & $\mathbf{11}$ & $8$ & $6$ & $8$ & $6$ & $\mathbf{11}$ & $9$ & $10$ \\ 
$TN(0.25,0.5)$ & $25$ & $\mathbf{28}$ & $26$ & $20$ & $9$ & $19$ & $12$ & $27$ & $23$ & $\mathbf{30}$ \\ 
$K(1,1.5)$ & $40$ & $\mathbf{47}$ & $46$ & $24$ & $12$ & $24$ & $17$ & $\mathbf{48}$ & $42$ & $\mathbf{47}$ \\ 
$K(1.5,1)$ & $40$ & $\mathbf{47}$ & $46$ & $24$ & $12$ & $24$ & $17$ & $\mathbf{48}$ & $42$ & $35$ \\ 
$K(1.5,2.5)$ & $58$ & $60$ & $58$ & $69$ & $20$ & $64$ & $34$ & $57$ & $\mathbf{74}$ & $\mathbf{81}$ \\ 
$\frac{3}{4}\beta(2,3)+\frac{1}{4}\mathcal{U}$ & $42$ & $41$ & $38$ & $\mathbf{60}$ & $16$ & $56$ & $25$ & $38$ & $47$ & $\mathbf{61}$ \\ 
$\frac{3}{4}\beta(1,0.5)+\frac{1}{4}\mathcal{U}$ & $59$ & $65$ & $76$ & $37$ & $33$ & $39$ & $35$ & $66$ & $\mathbf{80}$ & $\mathbf{77}$ \\ 
$\frac{3}{4}TN(0,0.5)+\frac{1}{4}\mathcal{U}$ & $46$ & $\mathbf{51}$ & $49$ & $30$ & $14$ & $30$ & $21$ & $\mathbf{52}$ & $39$ & $46$ \\ 
$\frac{3}{4}TN(0.25,0.25)+\frac{1}{4}\mathcal{U}$ & $79$ & $\mathbf{82}$ & $78$ & $70$ & $27$ & $67$ & $49$ & $80$ & $68$ & $\mathbf{85}$ \\ 
$\frac{3}{4}K(1.5,1)+\frac{1}{4}\mathcal{U}$ & $18$ & $\mathbf{20}$ & $19$ & $11$ & $7$ & $11$ & $9$ & $\mathbf{20}$ & $15$ & $19$ \\ 
$\frac{3}{4}K(2,1)+\frac{1}{4}\mathcal{U}$ & $33$ & $\mathbf{38}$ & $37$ & $20$ & $11$ & $20$ & $15$ & $\mathbf{38}$ & $28$ & $36$ \\ 
$S_1(0.7)$ & $33$ & $38$ & $\mathbf{45}$ & $19$ & $16$ & $19$ & $19$ & $39$ & $\mathbf{45}$ & $\mathbf{48}$ \\ 
$S_1(1.5)$ & $40$ & $\mathbf{47}$ & $46$ & $24$ & $12$ & $24$ & $17$ & $\mathbf{48}$ & $42$ & $\mathbf{47}$ \\ 
$S_2(0.7)$ & $14$ & $13$ & $28$ & $\mathbf{29}$ & $20$ & $27$ & $24$ & $12$ & $\mathbf{38}$ & $16$ \\ 
$S_2(1.5)$ & $8$ & $6$ & $6$ & $\mathbf{36}$ & $8$ & $\mathbf{33}$ & $7$ & $6$ & $22$ & $9$ \\ 
$S_3(0.7)$ & $10$ & $6$ & $5$ & $\mathbf{28}$ & $12$ & $\mathbf{27}$ & $12$ & $5$ & $10$ & $6$ \\ 
$S_3(1.5)$ & $19$ & $14$ & $20$ & $\mathbf{36}$ & $17$ & $33$ & $\mathbf{34}$ & $11$ & $13$ & $15$ \\ 
\hline \\[-1.8ex] 
\end{tabular} 
\end{table}

\begin{table}[t] \centering 
  \caption{Empirical powers associated with testing for the uniform distribution, $n=50$} 
  \label{Unif50} 
\begin{tabular}{@{\extracolsep{5pt}} lcccccccccc} 
\\[-1.8ex]\hline 
\hline \\[-1.8ex] 
Distributions & $KS_n$ & $CV_n$ & $AD_n$ & $W_n$ & $S_n$ & $K_n$ & $Q_n$ & $F_n$ & $ZC_n$ & $T_n$ \\
\hline
$\mathcal{U}$ & $5$ & $5$ & $5$ & $5$ & $5$ & $5$ & $5$ & $5$ & $5$ & $5$ \\ 
$\beta(2,2)$ & $23$ & $25$ & $37$ & $\mathbf{83}$ & $22$ & $\mathbf{79}$ & $29$ & $27$ & $82$ & $24$ \\ 
$\beta(2,3)$ & $97$ & $97$ & $99$ & $99$ & $56$ & $99$ & $88$ & $97$ & $\mathbf{100}$ & $\mathbf{100}$ \\ 
$\beta(1,0.5)$ & $96$ & $98$ & $\mathbf{99}$ & $84$ & $68$ & $87$ & $72$ & $98$ & $\mathbf{99}$ & $\mathbf{99}$ \\ 
$\beta(0.5,0.5)$ & $43$ & $46$ & $\mathbf{86}$ & $84$ & $58$ & $81$ & $61$ & $47$ & $\mathbf{95}$ & $43$ \\ 
$TN(0,0.5)$ & $92$ & $\mathbf{95}$ & $\mathbf{95}$ & $77$ & $34$ & $79$ & $59$ & $\mathbf{95}$ & $89$ & $\mathbf{95}$ \\ 
$TN(0,1)$ & $14$ & $\mathbf{16}$ & $15$ & $10$ & $6$ & $10$ & $7$ & $\mathbf{16}$ & $12$ & $15$ \\ 
$TN(0.25,0.5)$ & $40$ & $\mathbf{43}$ & $42$ & $31$ & $10$ & $29$ & $17$ & $\mathbf{43}$ & $36$ & $\mathbf{50}$ \\ 
$K(1,1.5)$ & $62$ & $70$ & $70$ & $39$ & $16$ & $41$ & $26$ & $\mathbf{71}$ & $66$ & $\mathbf{73}$ \\ 
$K(1.5,1)$ & $62$ & $\mathbf{70}$ & $\mathbf{70}$ & $39$ & $16$ & $41$ & $26$ & $\mathbf{71}$ & $66$ & $53$ \\ 
$K(1.5,2.5)$ & $86$ & $88$ & $90$ & $92$ & $33$ & $89$ & $65$ & $86$ & $\mathbf{97}$ & $\mathbf{97}$ \\ 
$\frac{3}{4}\beta(2,3)+\frac{1}{4}\mathcal{U}$ & $69$ & $69$ & $70$ & $\mathbf{85}$ & $25$ & $81$ & $44$ & $66$ & $74$ & $\mathbf{86}$ \\ 
$\frac{3}{4}\beta(1,0.5)+\frac{1}{4}\mathcal{U}$ & $81$ & $85$ & $\mathbf{92}$ & $58$ & $43$ & $61$ & $46$ & $86$ & $\mathbf{94}$ & $\mathbf{92}$ \\ 
$\frac{3}{4}TN(0,0.5)+\frac{1}{4}\mathcal{U}$ & $68$ & $\mathbf{73}$ & $72$ & $49$ & $17$ & $49$ & $30$ & $\mathbf{74}$ & $59$ & $70$ \\ 
$\frac{3}{4}TN(0.25,0.25)+\frac{1}{4}\mathcal{U}$ & $96$ & $\mathbf{97}$ & $96$ & $92$ & $40$ & $90$ & $70$ & $96$ & $90$ & $\mathbf{98}$ \\ 
$\frac{3}{4}K(1.5,1)+\frac{1}{4}\mathcal{U}$ & $27$ & $\mathbf{30}$ & $\mathbf{30}$ & $17$ & $8$ & $16$ & $11$ & $\mathbf{31}$ & $22$ & $27$ \\ 
$\frac{3}{4}K(2,1)+\frac{1}{4}\mathcal{U}$ & $51$ & $\mathbf{57}$ & $56$ & $32$ & $13$ & $33$ & $20$ & $\mathbf{57}$ & $43$ & $52$ \\ 
$S_1(0.7)$ & $50$ & $56$ & $\mathbf{64}$ & $30$ & $20$ & $31$ & $23$ & $58$ & $63$ & $\mathbf{67}$ \\ 
$S_1(1.5)$ & $62$ & $69$ & $\mathbf{70}$ & $39$ & $16$ & $41$ & $27$ & $\mathbf{70}$ & $66$ & $\mathbf{72}$ \\ 
$S_2(0.7)$ & $19$ & $18$ & $40$ & $\mathbf{45}$ & $23$ & $42$ & $28$ & $17$ & $\mathbf{53}$ & $22$ \\ 
$S_2(1.5)$ & $14$ & $12$ & $16$ & $\mathbf{58}$ & $12$ & $\mathbf{52}$ & $14$ & $12$ & $46$ & $15$ \\ 
$S_3(0.7)$ & $17$ & $11$ & $10$ & $\mathbf{45}$ & $16$ & $\mathbf{42}$ & $17$ & $8$ & $15$ & $9$ \\ 
$S_3(1.5)$ & $27$ & $21$ & $28$ & $\mathbf{58}$ & $21$ & $\mathbf{52}$ & $45$ & $16$ & $17$ & $20$ \\ 
\hline \\[-1.8ex] 
\end{tabular} 
\end{table}

The results in Tables \ref{Unif30} and \ref{Unif50} indicate that all of the tests considered maintain the specified nominal significance level of $5\%$. It is also evident that the newly developed test is competitive in terms of power; this test often outperforms the competing tests.

\begin{table}[t] \centering 
  \caption{Critical values for $T_n$ when testing for the normal distribution} 
  \label{CritvalsNorm} 
\begin{tabular}{@{\extracolsep{5pt}} lcccccccccc} 
\\[-1.8ex]\hline 
\hline \\[-1.8ex] 
$n$ & $10$ & $20$ & $30$ & $50$ & $75$ & $100$ & $150$ & $200$ & $300$ & $500$\\
\hline \\[-1.8ex] 
$\alpha=10\%$ & $0.052$ & $0.052$ & $0.053$ & $0.053$ & $0.053$ & $0.053$ & $0.053$ & $0.053$ & $0.053$ & $0.053$ \\ 
$\alpha=5\%$ & $0.064$ & $0.067$ & $0.067$ & $0.068$ & $0.068$ & $0.068$ & $0.068$ & $0.068$ & $0.068$ & $0.069$ \\ 
$\alpha=1\%$ & $0.096$ & $0.103$ & $0.104$ & $0.105$ & $0.106$ & $0.106$ & $0.106$ & $0.106$ & $0.106$ & $0.107$ \\ 
\hline \\[-1.8ex] 
\end{tabular} 
\end{table}

\begin{table}[t] \centering 
  \caption{Empirical powers associated with testing for the normal distribution, $n=30$} 
  \label{Normal30} 
\begin{tabular}{@{\extracolsep{5pt}} lcccccccccc} 
\\[-1.8ex]\hline 
\hline \\[-1.8ex] 
Distributions & $KS_n$ & $CV_n$ & $AD_n$ & $W_n$ & $S_n$ & $K_n$ & $Q_n$ & $F_n$ & $ZC_n$ & $T_n$ \\
\hline
$Z$ & $5$ & $5$ & $5$ & $5$ & $5$ & $5$ & $5$ & $5$ & $5$ & $5$ \\ 
$N(3,1)$ & $5$ & $5$ & $5$ & $5$ & $5$ & $5$ & $5$ & $5$ & $5$ & $5$ \\ 
$N(0,9)$ & $5$ & $5$ & $5$ & $5$ & $5$ & $5$ & $5$ & $5$ & $5$ & $5$ \\ 
$N(3,9)$ & $5$ & $5$ & $5$ & $5$ & $5$ & $5$ & $5$ & $5$ & $5$ & $5$ \\ 
$t(3)$ & $33$ & $39$ & $\mathbf{42}$ & $38$ & $20$ & $35$ & $22$ & $39$ & $\mathbf{46}$ & $36$ \\ 
$t(5)$ & $15$ & $18$ & $\mathbf{21}$ & $18$ & $10$ & $16$ & $10$ & $19$ & $\mathbf{26}$ & $18$ \\ 
$t(10)$ & $8$ & $8$ & $\mathbf{9}$ & $8$ & $7$ & $8$ & $6$ & $\mathbf{9}$ & $\mathbf{13}$ & $\mathbf{9}$ \\ 
$\frac{1}{2}Z+\frac{1}{2}N(1,4)$ & $16$ & $19$ & $\mathbf{21}$ & $17$ & $9$ & $15$ & $10$ & $20$ & $\mathbf{21}$ & $\mathbf{25}$ \\ 
$\frac{1}{2}Z+\frac{1}{2}N(1,9)$ & $35$ & $42$ & $\mathbf{43}$ & $42$ & $18$ & $37$ & $24$ & $42$ & $35$ & $\mathbf{46}$ \\ 
$\frac{1}{2}Z+\frac{1}{2}N(2,4)$ & $27$ & $32$ & $\mathbf{33}$ & $29$ & $11$ & $24$ & $14$ & $\mathbf{33}$ & $30$ & $\mathbf{34}$ \\ 
$SN(2)$ & $10$ & $11$ & $12$ & $10$ & $6$ & $9$ & $6$ & $11$ & $\mathbf{14}$ & $\mathbf{14}$ \\ 
$SN(2.5)$ & $13$ & $15$ & $17$ & $13$ & $7$ & $11$ & $8$ & $16$ & $\mathbf{19}$ & $\mathbf{19}$ \\ 
$SN(3)$ & $16$ & $19$ & $21$ & $17$ & $8$ & $14$ & $9$ & $20$ & $\mathbf{24}$ & $\mathbf{24}$ \\ 
$TN(-1,1)$ & $10$ & $15$ & $\mathbf{18}$ & $17$ & $9$ & $16$ & $10$ & $17$ & $\mathbf{20}$ & $17$ \\ 
$TN(-2,1)$ & $10$ & $13$ & $\mathbf{15}$ & $13$ & $8$ & $12$ & $8$ & $14$ & $14$ & $\mathbf{21}$ \\ 
$TN(-2.5,1.5)$ & $7$ & $7$ & $\mathbf{8}$ & $7$ & $5$ & $7$ & $6$ & $7$ & $6$ & $\mathbf{10}$ \\ 
$\chi^2(5)$ & $39$ & $50$ & $\mathbf{56}$ & $43$ & $21$ & $39$ & $19$ & $52$ & $\mathbf{64}$ & $\mathbf{56}$ \\ 
$\chi^2(10)$ & $21$ & $26$ & $30$ & $22$ & $11$ & $19$ & $10$ & $27$ & $\mathbf{36}$ & $\mathbf{33}$ \\ 
$\chi^2(15)$ & $16$ & $19$ & $21$ & $16$ & $8$ & $14$ & $8$ & $19$ & $\mathbf{26}$ & $\mathbf{24}$ \\ 
$\beta(2,1.5)$ & $10$ & $13$ & $\mathbf{15}$ & $14$ & $7$ & $13$ & $9$ & $14$ & $13$ & $\mathbf{20}$ \\ 
$\beta(2.5,2)$ & $7$ & $\mathbf{9}$ & $\mathbf{9}$ & $\mathbf{9}$ & $5$ & $\mathbf{9}$ & $7$ & $\mathbf{9}$ & $7$ & $\mathbf{12}$ \\ 
$\beta(3,2)$ & $9$ & $10$ & $\mathbf{11}$ & $\mathbf{11}$ & $6$ & $10$ & $7$ & $\mathbf{11}$ & $9$ & $\mathbf{15}$ \\ 
$HN(0,1)$ & $36$ & $49$ & $\mathbf{57}$ & $43$ & $27$ & $42$ & $21$ & $51$ & $\mathbf{67}$ & $49$ \\ 
\hline \\[-1.8ex] 
\end{tabular} 
\end{table} 

\begin{table}[!htbp] \centering 
  \caption{Empirical powers associated with testing for the normal distribution, $n=50$} 
  \label{Normal50} 
\begin{tabular}{@{\extracolsep{5pt}} lcccccccccc} 
\\[-1.8ex]\hline 
\hline \\[-1.8ex] 
Distributions & $KS_n$ & $CV_n$ & $AD_n$ & $W_n$ & $S_n$ & $K_n$ & $Q_n$ & $F_n$ & $ZC_n$ & $T_n$ \\
\hline
$Z$ & $5$ & $5$ & $5$ & $5$ & $5$ & $5$ & $5$ & $5$ & $5$ & $5$ \\ 
$N(3,1)$ & $5$ & $5$ & $5$ & $5$ & $5$ & $5$ & $5$ & $5$ & $5$ & $5$ \\ 
$N(0,9)$ & $5$ & $5$ & $5$ & $5$ & $5$ & $5$ & $5$ & $5$ & $5$ & $5$ \\ 
$N(3,9)$ & $5$ & $5$ & $5$ & $5$ & $5$ & $5$ & $5$ & $5$ & $5$ & $5$ \\ 
$t(3)$ & $47$ & $56$ & $\mathbf{59}$ & $55$ & $26$ & $52$ & $31$ & $56$ & $\mathbf{63}$ & $53$ \\ 
$t(5)$ & $20$ & $25$ & $\mathbf{29}$ & $25$ & $11$ & $23$ & $12$ & $26$ & $\mathbf{37}$ & $26$ \\ 
$t(10)$ & $8$ & $10$ & $\mathbf{11}$ & $10$ & $7$ & $9$ & $6$ & $10$ & $\mathbf{17}$ & $\mathbf{11}$ \\ 
$\frac{1}{2}Z+\frac{1}{2}N(1,4)$ & $24$ & $30$ & $\mathbf{32}$ & $27$ & $10$ & $23$ & $12$ & $31$ & $29$ & $\mathbf{39}$ \\ 
$\frac{1}{2}Z+\frac{1}{2}N(1,9)$ & $52$ & $63$ & $\mathbf{64}$ & $63$ & $23$ & $58$ & $33$ & $\mathbf{64}$ & $47$ & $\mathbf{68}$ \\ 
$\frac{1}{2}Z+\frac{1}{2}N(2,4)$ & $44$ & $53$ & $\mathbf{54}$ & $48$ & $14$ & $39$ & $19$ & $53$ & $46$ & $\mathbf{54}$ \\ 
$SN(2)$ & $13$ & $15$ & $17$ & $13$ & $7$ & $11$ & $7$ & $16$ & $\mathbf{19}$ & $\mathbf{20}$ \\ 
$SN(2.5)$ & $19$ & $23$ & $26$ & $20$ & $8$ & $16$ & $8$ & $24$ & $\mathbf{28}$ & $\mathbf{28}$ \\ 
$SN(3)$ & $25$ & $31$ & $34$ & $26$ & $10$ & $21$ & $10$ & $32$ & $\mathbf{38}$ & $\mathbf{36}$ \\ 
$TN(-1,1)$ & $15$ & $25$ & $\mathbf{34}$ & $29$ & $14$ & $26$ & $12$ & $29$ & $\mathbf{50}$ & $31$ \\ 
$TN(-2,1)$ & $16$ & $21$ & $27$ & $21$ & $11$ & $18$ & $11$ & $23$ & $\mathbf{34}$ & $\mathbf{38}$ \\ 
$TN(-2.5,1.5)$ & $8$ & $10$ & $11$ & $10$ & $7$ & $9$ & $7$ & $10$ & $\mathbf{12}$ & $\mathbf{16}$ \\ 
$\chi^2(5)$ & $59$ & $74$ & $\mathbf{80}$ & $66$ & $30$ & $62$ & $31$ & $75$ & $\mathbf{88}$ & $78$ \\ 
$\chi^2(10)$ & $34$ & $42$ & $48$ & $36$ & $14$ & $30$ & $13$ & $44$ & $\mathbf{58}$ & $\mathbf{50}$ \\ 
$\chi^2(15)$ & $24$ & $29$ & $34$ & $24$ & $10$ & $20$ & $10$ & $31$ & $\mathbf{41}$ & $\mathbf{37}$ \\ 
$\beta(2,1.5)$ & $15$ & $22$ & $27$ & $23$ & $10$ & $20$ & $10$ & $24$ & $\mathbf{32}$ & $\mathbf{37}$ \\ 
$\beta(2.5,2)$ & $10$ & $13$ & $\mathbf{15}$ & $14$ & $6$ & $13$ & $7$ & $14$ & $14$ & $\mathbf{21}$ \\ 
$\beta(3,2)$ & $12$ & $16$ & $\mathbf{19}$ & $16$ & $7$ & $14$ & $8$ & $17$ & $18$ & $\mathbf{27}$ \\ 
$HN(0,1)$ & $58$ & $75$ & $\mathbf{84}$ & $68$ & $42$ & $70$ & $39$ & $77$ & $\mathbf{93}$ & $73$ \\ 
\hline \\[-1.8ex] 
\end{tabular} 
\end{table}

Tables \ref{Normal30} and \ref{Normal50} show the results obtained for testing the hypothesis of normality. As was the case when testing for uniformity, the level of significance is maintained closely by all of the tests considered. The newly proposed test, $T_n$, is competitive in terms of power; $T_n$, $ZC_n$ and $AD_n$ provide the highest empirical powers for both sample sizes considered.

\begin{table}[!htbp] \centering 
  \caption{Critical values for $T_n$ when testing for the Pareto distribution} 
  \label{CritValsPareto} 
\begin{tabular}{@{\extracolsep{5pt}} lcccccccccc} 
\\[-1.8ex]\hline 
\hline \\[-1.8ex] 
$n$ & $10$ & $20$ & $30$ & $50$ & $75$ & $100$ & $150$ & $200$ & $300$ & $500$\\
\hline \\[-1.8ex] 
$\alpha=10\%$ &$0.058$ & $0.062$ & $0.064$ & $0.065$ & $0.066$ & $0.066$ & $0.066$ & $0.067$ & $0.067$ & $0.067$ \\ 
$\alpha=5\%$ &$0.074$ & $0.081$ & $0.083$ & $0.085$ & $0.086$ & $0.086$ & $0.087$ & $0.087$ & $0.087$ & $0.088$ \\ 
$\alpha=1\%$ & $0.116$ & $0.129$ & $0.132$ & $0.134$ & $0.136$ & $0.137$ & $0.138$ & $0.139$ & $0.138$ & $0.139$ \\ 
\hline \\[-1.8ex] 
\end{tabular} 
\end{table} 

\begin{table}[!htbp] \centering 
  \caption{Empirical powers associated with testing for the Pareto distribution, $n=30$} 
  \label{Pareto30} 
\begin{tabular}{@{\extracolsep{5pt}} lcccccccccc} 
\\[-1.8ex]\hline 
\hline \\[-1.8ex] 
Distributions & $KS_n$ & $CV_n$ & $AD_n$ & $W_n$ & $S_n$ & $K_n$ & $Q_n$ & $F_n$ & $ZC_n$ & $T_n$ \\
\hline
$\mbox{\mbox{P}}(0.5)$ & $5$ & $5$ & $5$ & $5$ & $5$ & $5$ & $5$ & $5$ & $5$ & $5$ \\ 
$\mbox{P}$ & $5$ & $5$ & $5$ & $5$ & $5$ & $5$ & $5$ & $5$ & $5$ & $5$ \\ 
$\mbox{P}(2)$ & $5$ & $5$ & $5$ & $5$ & $5$ & $5$ & $5$ & $5$ & $5$ & $5$ \\ 
$\mbox{P}(5)$ & $5$ & $5$ & $5$ & $5$ & $5$ & $5$ & $5$ & $5$ & $5$ & $5$ \\ 
$\Gamma(0.7)$ & $8$ & $8$ & $13$ & $11$ & $8$ & $11$ & $10$ & $9$ & $\mathbf{17}$ & $\mathbf{18}$ \\ 
$\Gamma(0.8)$ & $12$ & $14$ & $13$ & $13$ & $7$ & $13$ & $7$ & $\mathbf{15}$ & $\mathbf{15}$ & $\mathbf{19}$ \\ 
$\Gamma(1)$ & $35$ & $\mathbf{44}$ & $39$ & $35$ & $10$ & $34$ & $9$ & $\mathbf{46}$ & $39$ & $27$ \\ 
$W(0.7)$ & $8$ & $9$ & $16$ & $13$ & $9$ & $12$ & $12$ & $9$ & $\mathbf{20}$ & $\mathbf{18}$ \\ 
$W(0.8)$ & $10$ & $11$ & $11$ & $12$ & $7$ & $11$ & $7$ & $12$ & $\mathbf{13}$ & $\mathbf{18}$ \\ 
$W(0.9)$ & $20$ & $\mathbf{24}$ & $21$ & $20$ & $8$ & $19$ & $7$ & $\mathbf{25}$ & $21$ & $22$ \\ 
$LFR(0.2)$ & $45$ & $\mathbf{56}$ & $51$ & $46$ & $14$ & $46$ & $12$ & $\mathbf{59}$ & $51$ & $39$ \\ 
$LFR(0.5)$ & $55$ & $\mathbf{67}$ & $62$ & $57$ & $18$ & $56$ & $16$ & $\mathbf{69}$ & $61$ & $48$ \\ 
$LFR(1)$ & $63$ & $\mathbf{76}$ & $71$ & $66$ & $22$ & $66$ & $20$ & $\mathbf{78}$ & $69$ & $56$ \\ 
$EG(0.3)$ & $17$ & $\mathbf{21}$ & $17$ & $16$ & $6$ & $15$ & $5$ & $\mathbf{21}$ & $19$ & $14$ \\ 
$EG(0.4)$ & $12$ & $\mathbf{14}$ & $11$ & $11$ & $5$ & $11$ & $5$ & $\mathbf{14}$ & $13$ & $10$ \\ 
$\frac{3}{4}\Gamma(0.7)+\frac{1}{4}\mbox{P}$ & $13$ & $14$ & $17$ & $18$ & $13$ & $17$ & $15$ & $14$ & $\mathbf{20}$ & $\mathbf{22}$ \\ 
$\frac{3}{4}\Gamma(0.8)+\frac{1}{4}\mbox{P}$ & $19$ & $21$ & $19$ & $\mathbf{22}$ & $12$ & $\mathbf{22}$ & $12$ & $21$ & $16$ & $\mathbf{22}$ \\ 
$\frac{3}{4}\Gamma(1)+\frac{1}{4}\mbox{P}$ & $42$ & $\mathbf{50}$ & $44$ & $48$ & $16$ & $45$ & $15$ & $\mathbf{51}$ & $36$ & $28$ \\ 
$\frac{3}{4}W(0.7)+\frac{1}{4}\mbox{P}$ & $22$ & $21$ & $23$ & $\mathbf{26}$ & $14$ & $25$ & $18$ & $20$ & $22$ & $\mathbf{32}$ \\ 
$\frac{3}{4}W(0.8)+\frac{1}{4}\mbox{P}$ & $27$ & $\mathbf{28}$ & $24$ & $\mathbf{30}$ & $13$ & $\mathbf{28}$ & $14$ & $\mathbf{28}$ & $20$ & $\mathbf{28}$ \\ 
$\frac{3}{4}W(0.9)+\frac{1}{4}\mbox{P}$ & $34$ & $\mathbf{38}$ & $33$ & $\mathbf{38}$ & $13$ & $36$ & $14$ & $\mathbf{39}$ & $25$ & $28$ \\ 
$\frac{3}{4}LFR(0.2)+\frac{1}{4}\mbox{P}$ & $43$ & $\mathbf{52}$ & $46$ & $51$ & $17$ & $48$ & $17$ & $\mathbf{53}$ & $37$ & $29$ \\ 
$\frac{3}{4}LFR(0.5)+\frac{1}{4}\mbox{P}$ & $41$ & $\mathbf{50}$ & $45$ & $\mathbf{50}$ & $17$ & $47$ & $17$ & $\mathbf{52}$ & $36$ & $26$ \\ 
\hline \\[-1.8ex] 
\end{tabular} 
\end{table}

\begin{table}[!htbp] \centering 
  \caption{Empirical powers associated with testing for the Pareto distribution, $n=50$} 
  \label{Pareto50} 
\begin{tabular}{@{\extracolsep{5pt}} lcccccccccc} 
\\[-1.8ex]\hline 
\hline \\[-1.8ex] 
Distributions & $KS_n$ & $CV_n$ & $AD_n$ & $W_n$ & $S_n$ & $K_n$ & $Q_n$ & $F_n$ & $ZC_n$ & $T_n$ \\
\hline
$\mbox{P}(0.5)$ & $5$ & $5$ & $5$ & $5$ & $5$ & $5$ & $5$ & $5$ & $5$ & $5$ \\ 
$\mbox{P}$ & $5$ & $5$ & $5$ & $5$ & $5$ & $5$ & $5$ & $5$ & $5$ & $5$ \\ 
$\mbox{P}(2)$ & $5$ & $5$ & $5$ & $5$ & $5$ & $5$ & $5$ & $5$ & $5$ & $5$ \\ 
$\mbox{P}(5)$ & $5$ & $5$ & $5$ & $5$ & $5$ & $5$ & $5$ & $5$ & $5$ & $5$ \\ 
$\Gamma(0.7)$ & $10$ & $11$ & $17$ & $16$ & $9$ & $15$ & $11$ & $12$ & $\mathbf{24}$ & $\mathbf{30}$ \\ 
$\Gamma(0.8)$ & $19$ & $\mathbf{22}$ & $20$ & $19$ & $8$ & $18$ & $8$ & $\mathbf{22}$ & $\mathbf{22}$ & $\mathbf{35}$ \\ 
$\Gamma(1)$ & $55$ & $\mathbf{68}$ & $64$ & $55$ & $15$ & $54$ & $15$ & $\mathbf{70}$ & $62$ & $53$ \\ 
$W(0.7)$ & $10$ & $12$ & $22$ & $19$ & $10$ & $17$ & $13$ & $13$ & $\mathbf{28}$ & $\mathbf{29}$ \\ 
$W(0.8)$ & $15$ & $17$ & $17$ & $17$ & $8$ & $16$ & $8$ & $18$ & $\mathbf{19}$ & $\mathbf{32}$ \\ 
$W(0.9)$ & $32$ & $39$ & $35$ & $31$ & $10$ & $30$ & $10$ & $\mathbf{41}$ & $34$ & $\mathbf{40}$ \\ 
$LFR(0.2)$ & $69$ & $\mathbf{81}$ & $\mathbf{78}$ & $70$ & $21$ & $70$ & $23$ & $83$ & $77$ & $69$ \\ 
$LFR(0.5)$ & $78$ & $\mathbf{89}$ & $87$ & $81$ & $27$ & $81$ & $31$ & $\mathbf{91}$ & $86$ & $79$ \\ 
$LFR(1)$ & $86$ & $\mathbf{94}$ & $93$ & $88$ & $34$ & $89$ & $40$ & $\mathbf{95}$ & $92$ & $86$ \\ 
$EG(0.3)$ & $27$ & $\mathbf{33}$ & $29$ & $24$ & $8$ & $24$ & $7$ & $\mathbf{34}$ & $31$ & $26$ \\ 
$EG(0.4)$ & $17$ & $\mathbf{21}$ & $18$ & $15$ & $6$ & $15$ & $5$ & $\mathbf{21}$ & $20$ & $18$ \\ 
$\frac{3}{4}\Gamma(0.7)+\frac{1}{4}P$ & $19$ & $19$ & $24$ & $\mathbf{26}$ & $15$ & $25$ & $19$ & $19$ & $\mathbf{26}$ & $\mathbf{29}$ \\ 
$\frac{3}{4}\Gamma(0.8)+\frac{1}{4}\mbox{P}$ & $28$ & $31$ & $27$ & $\mathbf{33}$ & $14$ & $\mathbf{32}$ & $16$ & $31$ & $21$ & $31$ \\ 
$\frac{3}{4}\Gamma(1)+\frac{1}{4}\mbox{P}$ & $62$ & $\mathbf{70}$ & $66$ & $69$ & $21$ & $66$ & $25$ & $\mathbf{72}$ & $50$ & $45$ \\ 
$\frac{3}{4}W(0.7)+\frac{1}{4}\mbox{P}$ & $34$ & $33$ & $35$ & $\mathbf{41}$ & $18$ & $39$ & $24$ & $31$ & $31$ & $\mathbf{48}$ \\ 
$\frac{3}{4}W(0.8)+\frac{1}{4}\mbox{P}$ & $42$ & $43$ & $38$ & $\mathbf{45}$ & $16$ & $43$ & $20$ & $42$ & $27$ & $\mathbf{44}$ \\ 
$\frac{3}{4}W(0.9)+\frac{1}{4}\mbox{P}$ & $51$ & $\mathbf{57}$ & $50$ & $56$ & $17$ & $53$ & $20$ & $\mathbf{57}$ & $35$ & $42$ \\ 
$\frac{3}{4}LFR(0.2)+\frac{1}{4}\mbox{P}$ & $62$ & $\mathbf{72}$ & $68$ & $\mathbf{72}$ & $23$ & $70$ & $28$ & $\mathbf{74}$ & $51$ & $45$ \\ 
$\frac{3}{4}LFR(0.5)+\frac{1}{4}\mbox{P}$ & $59$ & $70$ & $66$ & $\mathbf{71}$ & $23$ & $69$ & $29$ & $\mathbf{72}$ & $49$ & $41$ \\ 
\hline \\[-1.8ex] 
\end{tabular} 
\end{table}

In addition to indicating that the tests for the Pareto distribution maintain the nominal significance level, Tables \ref{Pareto30} and \ref{Pareto50} show that the power performance of $T_n$ compares well to those of the remaining tests considered.

\section{An application in financial modelling}

In financial modelling the expected payoff of options, and therefore the prices of options, are functions of the distribution of the log-returns of the underlying stock. The famous Black-Merton-Scholes model calculates option prices based on the assumption that observed financial log-returns are realised from a normal distribution. This assumption has been challenged and criticised in the financial literature; see, for example, \cite{Cont}. In order to demonstrate the practical use of the tests considered, we test the hypothesis that observed financial log-returns are realised from a normal distribution.

The Standard \& Poor 500 index is a market-capitalisation weighted index consisting of 500 large companies in the United States. The data considered are 50 consecutive daily log-returns of the Standard \& Poor 500 index for the period ending on 2 August 2021. The relevant index prices can be downloaded from \url{http://finance.yahoo.com}.

Table 10 shows the $p$-values associated with each of the tests discussed in Section \ref{sec:SimuUn} obtained using one million bootstrap samples in each case. Based on the $p$-values in the table, none of the tests rejects the hypothesis of normality for any reasonable nominal significance level. Therefore, we conclude that the normality assumption implicit in the Black-Merton-Scholes model will lead to realistic option prices in this case.

\begin{table}[!htbp] \centering 
  \caption{$p$-values associated with the various tests considered} 
  \label{} 
\begin{tabular}{@{\extracolsep{5pt}} lcccccccccc} 
\\[-1.8ex]\hline 
\hline \\[-1.8ex] 
Test & $KS_n$ & $CV_n$ & $AD_n$ & $WA_n$ & $S_n$ & $K_n$ & $QM_n$ & $F_n$ & $ZC_n$ & $U_n$ \\
$p$-value & $0.288$ & $0.178$ & $0.203$ & $0.147$ & $0.346$ & $0.109$ & $0.571$ & $0.170$ & $0.557$ & $0.194$\\
\hline \\[-1.8ex] 
\end{tabular} 
\end{table} 

%

\section{Conclusion and Outlook}

In this paper, we propose a new omnibus test of fit for univariate distributions based on a characterisation of the uniform distribution. The asymptotic properties of this measure are considered for the case when a simple hypothesis is tested. The finite sample power performance of the newly proposed test is compared to that of other omnibus tests not containing a tuning parameter in both the simple and composite hypothesis cases. In both cases we demonstrate that the test is competitive in terms of power performance, often outperforming the classical goodness-of-fit tests. A practical example relating to the use of the test in finance is also provided.

We conclude this paper by pointing out some open questions for further research. Regarding the results of Theorem \ref{thm:asydistsh} it is well known from theory of Gaussian processes that by the orthogonal decomposition of the process $Z$, see \cite{SW:1986} p. 206, $\|Z\|^2_{\mathbb{H}}=\sum_{j=1}^\infty \lambda_j N_j^2,$ 
where $N_1,N_2,\ldots$ are i.i.d. standard normal, and $\lambda_1,\lambda_2,\ldots$ is a positive decreasing sequence of eigenvalues of the integral operator $
\mathcal{K}:\mathbb{H}\rightarrow\mathbb{H},\, f\mapsto\mathcal{K}f(\cdot)=\int_0^1K_Z(\cdot,t)f(t)\mbox{d}t.$
To calculate the eigenvalues $\lambda_j$, $j=1,2,\ldots$, of $\mathcal{K}$, one has to solve the homogeneous Fredholm integral equation of the second kind;
\begin{equation*}\label{eq:inteq}
\int_0^1 K_Z(x,t)f(t)\mbox{d}t=\lambda f(x),\quad 0< x< 1,
\end{equation*}
see, for example, \cite{KS:1947}. The largest eigenvalue, $\lambda_1$, is used to approximate the local Bahadur slope, which combined with the corresponding $\Delta$ in \eqref{eq:aslimit} for an alternative distribution leads to approximate Bahadur efficiencies; see \cite{B:1960} and \cite{N:1995}. Unfortunately we did not succeed in calculating explicit values of $\lambda_1$ due to the complexity of the problem, but approximation methods such as those presented in Section 5 in \cite{EH:2021} may be applied. 

Interesting generalisations of the testing problem in $\eqref{eq:H0unif}$ include testing for the uniform distribution with unknown support. This testing problem is considered in \cite{BH:1990}. A start for this generalisation is the adaptation of the characterisation of the 4-parameter beta distribution in Theorem 1 of \cite{A:1991} together with the derivation of a test statistic similar to $T_n$.

Testing the assumption of multivariate uniformity has received less research attention, but recently tests for the uniform distribution with known support have been proposed based on random geometric graphs, see \cite{ENS2020}, and on distance to boundary methods, see \cite{BCV:2006}. The case of unknown support is treated in \cite{BCP:2012a,BCP:2012b}. None of these methods rely on a characterisation of multivariate uniformity, hence extending the test studied in this paper would be of interest in this case.

Other possible directions for further research include deriving the asymptotic theory of the proposed tests of normality and for the Pareto distribution as well as studying  the performance of the test proposed above when testing for other classes of distributions not considered in this paper. As specific examples, we mention the exponential distribution (for a review of testing procedures for this distribution, see \cite{allison2017apples}) and the Rayleigh distribution (a recent reference is \cite{LNA2020}). Finally, we mention that the test proposed in this paper may be generalised to test a goodness-of-fit assumption in the presence of random right censoring; for recent references related to testing the assumption that data are realised from the exponential distribution, see \cite{bothma2020kaplan,BM}.

\bibliographystyle{plain}
\bibliography{lit-AC}  

\end{document}